\documentclass[preprint,12pt]{elsarticle}

\usepackage{amssymb}

\usepackage{natbib}

\usepackage{bstnotations}

\newcommand{\veZ}{\ve{Z}}
\newcommand{\vez}{\ve{z}}
\newcommand{\veX}{\ve{X}}
\newcommand{\vex}{\ve{x}}

\usepackage{mathtools}

\usepackage[bookmarksopen=true]{hyperref}

\journal{Reliability Engineering \& System Safety}

\begin{document}

\begin{frontmatter}



\title{Computing derivative-based global sensitivity measures using polynomial chaos expansions}


%

\author[1]{B. Sudret \corref{cor1}}
\ead{sudret@ibk.baug.ethz.ch}

\ead[url]{http://www.ibk.ethz.ch/su/people/sudretb}

\author[1]{C.V. Mai}
\ead{mai@ibk.baug.ethz.ch}
\cortext[cor1]{Corresponding author}

\address[1]{ETH Z\"{u}rich, Institute of Structural Engineering,
Chair of Risk, Safety \& Uncertainty Quantification, Stefano-Franscini-Platz 5, CH-8093 Z\"{u}rich, Switzerland}

\begin{abstract}
  In the field of computer experiments sensitivity analysis aims at
  quantifying the relative importance of each input parameter (or
  combinations thereof) of a computational model with respect to the
  model output uncertainty. Variance decomposition methods leading to
  the well-known Sobol' indices are recognized as accurate techniques,
  at a rather high computational cost though. The use of polynomial
  chaos expansions (PCE) to compute Sobol' indices has allowed to
  alleviate the computational burden though. However, when dealing with
  large dimensional input vectors, it is good practice to first use
  screening methods in order to discard unimportant variables.  The {\em
    derivative-based global sensitivity measures} (DGSM) have been
  developed recently in this respect. In this paper we show how
  polynomial chaos expansions may be used to compute analytically DGSMs
  as a mere post-processing. This requires the analytical derivation of
  derivatives of the orthonormal polynomials which enter PC expansions.
  The efficiency of the approach is illustrated on two well-known
  benchmark problems in sensitivity analysis.
\end{abstract}


\end{frontmatter}



\section{Introduction}
\label{sec01}

Nowadays, the increasing computing power allows one to use numerical
models to simulate or predict the behavior of physical systems in
various fields, \eg mechanical engineering~\citep{Reuter2008}, civil
engineering~\citep{Hasofer2009}, chemistry~\citep{Campolongo2007}, etc.
The considered systems usually lead to highly complex models with
numerous input factors (possibly tens to
hundreds~\citep{SudretMaiIcossar2013, Patelli2010a}) that are required to
represent all the parameters driving the system's behaviour, \eg
boundary and initial conditions, material properties, external
excitations, etc.  On the one hand, this increases dramatically the
computational cost.  On the other hand, the input factors that are not
perfectly known may introduce uncertainties into the predictions of the
system response.  In order to take into account the uncertainty,
probabilistic approaches have been developed in the last two decades, in
which the model input parameters are represented by random variables.
Then the input uncertainties are propagated through the computational
model and the distribution, moments or probability of exceeding
prescribed thresholds may be computed \citep{Sudret2007,Derocquigny2012}.

In this context, sensitivity analysis (SA) examines the sensitivity of
the model output with respect to the input parameters, \ie how the
output variability is affected by the uncertain input
factors~\citep{Saltelli2000, Saltelli2004,Saltelli2008}.  The use of SA
is common in various fields: engineering~\citep{Hasofer2009,
  Pappenberger2008, Reuter2008, Kala2011},
chemistry~\cite{Campolongo2007}, nuclear safety~\cite{Fasso2013},
economy~\cite{Borgonovo2006}, biology~\cite{Marino2008}, and
medicine~\cite{Abraham2007}, among others.  One can traditionally
classify SA into \emph{local} and \emph{global} sensitivity
analyses.  The former aims at assessing the output
sensitivity to small input perturbations around the nominal values of
input parameters.  The latter aims at assessing the overall or average
influence of input parameters onto the output.  Local SA has the
disadvantages of being related to a fixed nominal point in the input
space, and the interaction between the inputs is not accounted
for~\cite{Kucherenko2009}.  On the other hand, global SA techniques take
into account the input interaction and are not based on the choice of a
nominal point but account for the whole input space.

The most common sensitivity analysis methods found in the literature are
the method of \cite{Morris1991}, FAST
\cite{Cukier1973,Cukier1978,Mara2009} and variance decomposition
methods originally investigated in \cite{Sobol1993,
  Sobol2001,Sobol2005, Archer1997,Saltelli2002, Saltelli2010}. Usually
standard Monte Carlo simulation (MCS) is employed for estimating the
sensitivity indices in all these techniques.  This approach requires a
large number of model evaluations though and therefore is usually
computationally expensive.  When complex systems are of interest, the
standard MCS becomes unaffordable.  To overcome this problem, metamodels
(also called {\em surrogate models} or emulators) are usually used in
order to carry out the Monte Carlo simulation
\cite{Sathyanarayanamurthy2009,Zhang2014}.  In particular, polynomial
chaos expansions (PCE) have been recognized as a versatile tool for
building surrogate models and for conducting reliability and sensitivity
analyses, as originally shown in \cite{SudretCSM2006,
  SudretRESS2008b,BlatmanRESS2010}. Using PCE, variance-based
sensitivity analysis becomes a mere post-processing of the polynomial
coefficients once they have been computed.

More recently, a new gradient-based technique has been proposed for
screening unimportant factors. The so-called {\em derivative-based
  global sensitivity measures} (DGSM) are shown to be upper bounds of
the total Sobol' indices while being less computationally demanding
\cite{Sobol2009, Sobol2010, Kucherenko2009, Lamboni2013}.  Although the
computational cost of this technique is reduced compared to the
variance-based technique \cite{Kucherenko2009}, its practical computation
still relies on the Monte Carlo simulation approach.

In this paper we investigate the potential of polynomial chaos
expansions for computing derivative-based sensitivity indices and allow
for an efficient screening procedure. The paper is organized as follows:
the classical derivation of Sobol' indices and their link to
derivative-based sensitivity indices is summarized in
Section~\ref{sec02}. The machinery of polynomial chaos expansions and
the link with sensitivity analysis is developed in
Section~\ref{sec03}. The computation of the DGSM based on PC expansions
is then presented in Section~\ref{sec04}, in which an original method
for computing the derivatives of orthogonal polynomials is
presented. Finally two numerical tests are carried out in
Section~\ref{sec05}.

\section{Derivative-based global sensitivity measures}

\label{sec02}

\subsection{Variance-based sensitivity measures}
\label{sec02.1} Global sensitivity analysis (SA) aims at quantifying the
impact of input parameters onto the output quantities of interest.  One
input factor is considered insignificant (unessential) when it has
little or no effect on the output variability.  In practice, screening
out the insignificant factors allows one to reduce the dimension of the
problem, \eg by fixing the unessential parameters.

\emph{Variance-based} SA relies upon the decomposition of the output
variance into contributions of different components, \ie marginal
effects and interactions of input factors.  Consider a numerical model
$Y=\cm(\ve{X})$ where the input vector $\veX$ contains $M$ independent
input variables $\veX=\acc{X_1 \enum X_M}$ with uniform distribution
over the unit-hypercube $\ch^M$ and $Y$ is the scalar output.  The
\emph{Sobol' decomposition} reads~\cite{Sobol1993}:
\begin{equation}
\begin{split} Y =\cm(\ve{X})= & \cm_0 + \sum\limits_{i=1}^M \cm_i (X_i)
+ \sum\limits_{1 \leq i < j \leq M} \cm_{i,j}(X_i,X_j) + \dots \\ & +
\cm_{1 \enum M}(X_1\enum X_M)
\end{split}
\label{eq01}
\end{equation} 
in which $\cm_0 = \Esp{\cm(\veX)}$ is a constant term and
each summand $\cm_{i_1 \enum i_s} ( X_{i_1} \allowbreak \enum X_{i_s} )$
is a function of the variables $\acc{ X_{i_1} \enum X_{i_s}}, s \leq M$.
For the sake of conciseness we introduce the following notation for the
subset of indices:
\begin{equation} 
\iu \eqdef \acc{i_1 \enum i_s}
\end{equation} and denote by $\veX_{\iu}$ the subvector of $\ve{X}$ that
consists of the variables indexed by  $\iu$.  Using this set notation,
\eqrefe{eq01} rewrites:
\begin{equation} 
Y \eqdef \cm_0 + \sum_{\begin{subarray}{c}\iu \subset
\acc{1 \enum M} \\ \iu \neq \bs{0} \end{subarray} }
\cm_{\iu}(\ve{X}_{\iu}),
\end{equation} 
in which $\cm_{\iu}(\ve{X}_{\iu})$ is the summand
including the subset of parameters $\ve{X}_{\iu}$.  According
to~\cite{Sobol1993}, a unique decomposition requires the orthogonality
of the summands, \emph{i.e.}:
\begin{equation} 
\Esp{\cm_{\iu} (\ve{X}_{\iu}) \, \cm_{\iv}
(\ve{X}_{\iv})} = \int_{ \ch^M} \cm_{\iu} (\ve{x}_{\iu}) \, \cm_{\iv}
(\ve{x}_{\iv}) \, \di \ve{x} = 0 \quad , \quad \iu \neq \iv
\end{equation} 
In particular each summand shall be of zero mean value.
Accordingly the variance of the response $Y=\cm (\ve{X})$ reads:
\begin{equation} D \eqdef \Var{Y}=\sum\limits_{\begin{subarray}{c}\iu
\subset \acc{1 \enum M} \\ \iu \neq \bs{0} \end{subarray} }
\Var{\cm_{\iu}(\ve{X}_{\iu}) }.
 \label{eq02}
\end{equation} In this expansion $\Var{\cm_{\iu}(\ve{X}_{\iu}) }$ is the
contribution of summand $\cm(\veX_{\iu})$ to the output variance.

The \emph{Sobol' sensitivity index} $S_{\iu}$ for the subset of
variables $\veX_{\iu}$ is defined as follows~\cite{Sobol2001}:
  \begin{equation} S_{\iu} \eqdef \dfrac{D_{\iu}}{D} = \dfrac{
\Var{\cm_{\iu}(\ve{X}_{\iu})}}{D}
   \label{eq03}
  \end{equation}
The \emph{total sensitivity index} for subset $\veX_{\iu}$ is given
by~\cite{Sobol2001}:
  \begin{equation} S_{\iu}^T \eqdef \dfrac{D_{\iu}^T}{D} = \sum\limits_{
\iv \supset \iu} \dfrac{ \Var{\cm_{\iv}(\ve{X}_{\iv})}}{D}
   \label{eq04}
 \end{equation} 
 where the sum is extended over all sets $\iv=\acc{j_1 \enum j_t}$ which
 contains $\iu$.  It represents the total amount of uncertainty
 apportioned to the subset of variables $\veX_{\iu}$.  For instance, for
 a single variable $X_i, \, i=1 \enum M$ the Sobol' sensitivity index
 reads:
\begin{equation} 
S_i=\dfrac{\Var{\cm_i(X_i)}}{D},
\label{eq05}
\end{equation}
and the total Sobol' sensitivity index reads:
\begin{equation} S_i^T = \sum\limits_{\iv \ni i} \dfrac{
\Var{\cm_{\iv}(\ve{X}_{\iv})}}{D}.
 \label{eq05b}
\end{equation} $S_i$ and $S_i^T$ respectively represent the sole and
total effect of the factor $X_i$ on the system's output variability.
The smaller $S_i^T$ is, the less important the factor $X_i$ is. In the
case when $S_i^T \ll 1$, $X_i$ is considered as unimportant (unessential
or insignificant) and may be replaced in the analysis by a deterministic
value.

In the literature one can find different approaches for computing the
total Sobol' indices, such as the Monte Carlo simulation (MCS) and the
spectral approach.  \cite{Sobol1993, Sobol2001} proposed direct
estimation of the sensitivity indices for subsets of variables using
only the model evaluations at specially selected points.  The approach
relies on computing analytically the integral representations of
$D_{\iu}$ and $D_{\iu}^T$ respectively defined in \eqrefe{eq03} and
\eqrefe{eq04}.

Let us denote by $\overline{\iu}$ the set that is complementary to
$\iu$, \ie $\veX=(\veX_{\iu},\veX_{\overline{\iu}})$.  Let $\veX$ and
$\veX'$ be vectors of independent uniform variables defined on the unit
hypercube $\ch^M$ and define $\veX'=(\veX_{\iu}^{'},
\veX_{\overline{\iu}}^{'})$. The partial variance $D_{\iu}$ is
represented as follows~\cite{Sobol2005}:
\begin{equation} 
D_{\iu} = \iint\cm(\ve{x})
\cm(\vex_{\iu},\vex_{\overline{\iu}}^{'}) \, \di \vex \;\di
\vex_{\overline{\iu}}^{'} - {\cm_0}^2
\end{equation} 
The total variance $D_{\iu}^T$ is given by~\cite{Sobol2005}:
\begin{equation} 
D_{\iu}^T = \dfrac{1}{2} \iint \bra{
\cm(\vex) - \cm(\vex_{\iu}^{'},\vex_{\overline{\iu}}) } ^2 \, \di
\vex \;\di \vex_{\overline{\iu}}
\end{equation} 
A Monte Carlo algorithm is used to estimate the above
integrals.  For each sample point, one generates two $M$-dimensional
samples $\ve{x}=({\ve{x}_{\iu}},{\ve{x}_{\overline{\iu}}})$ and
$\ve{x}^{'}=({\ve{x}_{\iu}}^{'},{\ve{x}_{\overline{\iu}}}^{'})$.  The
function is evaluated at three points
$({\ve{x}_{\iu}},{\ve{x}_{\overline{\iu}}})$,
$(\ve{x}_{\iu}^{'},{\ve{x}_{\overline{\iu}}})$ and
$({\ve{x}_{\iu}},\ve{x}_{\overline{\iu}}^{'})$.  Using $N$ independent
sample points, one computes the quantities of interest $D$, $D_{\iu}$
and $D_{\iu}^T$ by means of the following crude Monte Carlo estimators:
\begin{equation} \cm_0= \dfrac{1}{N} \sum\limits_{i=1}^{N}
\cm\prt{\ve{x}^{(i)}}
\end{equation}
\begin{equation} D + {\cm_0}^2= \dfrac{1}{N} \sum\limits_{i=1}^{N}
\bra{\cm\prt{\ve{x}^{(i)}}}^2
\end{equation}
\begin{equation} D_{\iu} + {\cm_0}^2= \dfrac{1}{N} \sum\limits_{i=1}^{N}
\cm\prt{\ve{x}^{(i)}}
\cm\prt{{\ve{x}_{\iu}^{(i)}},{\ve{x}_{\overline{\iu}}^{{(i)}'}}}
\end{equation}
\begin{equation} D_{\iu}^T = \dfrac{1}{N} \sum\limits_{i=1}^{N}
\dfrac{1}{2} \bra{ \cm\prt{\ve{x}^{(i)}} -
\cm\prt{{\ve{x}_{\iu}^{{(i)}'}},{\ve{x}_{\overline{\iu}}^{(i)}}} } ^2
\end{equation}
The computation of the sensitivity indices by MCS may exhibit reduced
accuracy when the mean value $\cm_0$ is large.  In addition, the
computational cost is prohibitive: in order to compute the sensitivity
indices for $M$ parameters, MCS requires $(M+2)\times N$ model runs, in
which $N$ is the number of sample points typically chosen equal to
$10^3-10^4$ to reach an acceptable accuracy.  \cite{Saltelli2002}
suggested a procedure that is more efficient for computing the first and
total sensitivity indices.  \cite{Sobol2007} modified the MCS procedure
in order to reduce the lack of accuracy.  Some other estimators for the
sensitivity indices by MCS may be found in \cite{Monod2006,Janon2013}.

\subsection{Derivative-based sensitivity indices} 
The total Sobol' indices may be used for screening purposes. Indeeed a
negligible total Sobol' index $S_i^T$ means that variable $X_i$ does not
contribute to the output variance, neither directly nor in interaction
with orther variables. In order to avoid the computational burden
associated with estimating all total Sobol' indices, a new technique
based on derivatives has been recently proposed by
\cite{Kucherenko2009}.

Derivative-based sensitivity analysis originates from the \emph{Morris
  method} introduced in~\cite{Morris1991}. The idea is to measure the
average of the \emph{elementary effects} over the input space.
Considering variable $X_i$ one first samples an experimental design (ED)
in the input space $\cx = \acc{\vex^{(1)} \enum \vex^{(N)}}$ and then
varies this sample in the $i^{th}$ direction. The elementary effect
($EE$) is defined as:
\begin{equation} EE_i^{(j)}=\dfrac{\cm(\vex_r^{(j)}) -
\cm(\vex^{(j)})}{\Delta}
 \label{eq06}
\end{equation}
in which $\vex^{(j)}=\acc{ x_1^{(j)} \enum x_i^{(j)} \enum x_M^{(j)} }$
is the $j^{th}$ sample point and $\vex_r^{(j)} = \acc{ x_1^{(j)} \enum
  x_i^{(j)}+\Delta \enum x_M^{(j)} }$ is the perturbed sample point in
the $i$-th direction.  The Morris importance measure (Morris factor) is
defined as the average of the $EE_i$'s:
\begin{equation} 
  \mu_i= \dfrac{1}{N} \sum\limits_{j=1}^{N} EE_i^{(j)}
 \label{eq07}
\end{equation}
By definition, the variance $\sigma_i^2$ of the $EE$s is calculated
from:
\begin{equation} \sigma_i^2 = \dfrac{1}{N-1} \sum\limits_{j=1}^{N}
(EE_i^{(j)} - \mu_i)^2
 \label{eq08}
\end{equation}
Kucherenko \etal \citep{Kucherenko2009} generalized these quantities as
follows:
\begin{equation} 
  \mu_i \eqdef\Esp{\dfrac{\partial \cm}{\partial x_i} (\veX) }
  = \int_{\ch^M} \dfrac{\partial \cm}{\partial x_i} (\ve{x}) \di \vex
 \label{eq09}
\end{equation}
\begin{equation} 
\sigma_i^2 = \int\limits_{\ch^M} \bra{\dfrac{\partial
\cm}{\partial x_i} (\vex) }^2 \, \di \vex - \mu_i^2
 \label{eq10}
\end{equation}
provided that $\dfrac{\partial \cm}{\partial x_i}$ is square-integrable.
Any input parameter $X_i$ with $\mu_i\ll 1$ \emph{and} $\sigma_i \ll 1$
is considered as unimportant. It has been shown that the Morris factor
has a higher convergence rate compared to variance-based methods, which
makes it attractive from a computational viewpoint.
Because the elementary effects may be positive or negative, they can
cancel each other, which might lead to a misinterpretation of the
importance of $X_i$.  To avoid this,~\citet{Campolongo2007} modified
the Morris factor as follows:
\begin{equation} \mu_i^{\ast} =\Esp{\abs{\dfrac{\partial \cm}{\partial
x_i} (\veX) }}
 \label{eq11}
\end{equation}

Recently,~\citet{Sobol2009} introduced a new sensitivity measure (SM)
which is the mean-squared derivative of the model with respect
to $X_i$:
\begin{equation} \nu_i=\Esp{\prt{\dfrac{\partial \cm}{\partial x_i}
(\veX) }^2}
 \label{eq12}
\end{equation} \citet{Sobol2009} and \citet{ Lamboni2013} could
establish a link between the $\nu_i$ in~\eqrefe{eq12} and the total
Sobol' indices in \eqrefe{eq05b}.  In case $X_i$ is a uniform random
variable over $[0,\,1]$, one gets:
\begin{equation} S_i^T \leq S_i^{DGSM} \eqdef \dfrac{\nu_i}{\pi^2 D}
\end{equation} where $S_i^{DGSM}$ is the upper-bound to the total
sensitivity index $S_i^T$ and $D$ is the model output variance.  In case
of a uniform variable $X_i \sim [a_i,\,b_i]$ this upper bound scales to:
\begin{equation} S_i^T \leq S_i^{DGSM} = \dfrac{(b_i-a_i)^2}{\pi^2}
\dfrac{\nu_i}{D}
  \label{eq13}
\end{equation} Finally the above results can be extended to other types
of distributions.  If $X_i \sim \cn \prt{a_i, b_i}$ is a Gaussian random
variable with mean and variance $a_i$ and $b_i^2$ respectively, one gets:
\begin{equation} S_i^T \leq S_i^{DGSM} = {b_i}^2 \dfrac{\nu_i}{D}
 \label{eq14}
\end{equation}
In the general case,~\citet{Lamboni2013} define the upper bound of the
total Sobol' index of $X_i$ as:
\begin{equation} 
S_i^{DGSM} =4 {C_i}^2 \dfrac{\nu_i}{D}
\end{equation} 
in which $C_i= \sup \limits_{\substack{x\in \Rr}}
\dfrac{\min{\bra{F_{X_i}(x),1-F_{X_i}(x)}}}{f_{X_i}(x)}$ is the Cheeger
constant, $F_{X_i}$ is the cumulative distribution function of $X_i$ and
$f_{X_i}$ is the probability density function of $X_i$.


\section{Polynomial chaos expansions}
\label{sec03}

Let us consider a numerical model $Y=\cm(\ve{X})$ where the input vector
$\veX$ is composed of $M$ \emph{independent} random variables $\veX=
\acc{X_i,\, i=1 \enum M}$ and $Y$ is the output quantity of interest.
Assuming that $Y$ has a finite variance, it can be represented as
follows \cite{Ghanembook2003, Soize2004}:
\begin{equation} Y= \cm(\veX)= \sum\limits_{j=0}^{\infty} y_j
\phi_j(\veX)
 \label{eq3.1}
\end{equation} in which $\acc{ \phi_j(\veX), j=0 \enum \infty }$ form a
basis on the space of second order random variables and $y_j$'s are the
coordinates of $Y$ onto this basis.  In case the basis terms are
multivariate orthonormal polynomials of the input variables $\veX$,
\eqrefe{eq3.1} is called polynomial chaos expansion.

Assuming that the input vector $\veX$ has independent components $X_i$
with prescribed probability distribution functions $f_{X_i}$, one
obtains the joint probability density function:
\begin{equation} f_{\veX}(\vex)= \prod\limits_{i=1}^M f_{X_i}(x_i)
 \label{eq3.2}
\end{equation} For each $X_i$, one can construct a family of orthogonal
univariate polynomials $\acc{ P_k^{(i)}, k \in \Nn }$ with respect to
the probability measure $\Pp_{X_i}(\di x_i)=f_{X_i}(x_i) \di x_i$
satifying:
\begin{equation} \langle P_j^{(i)}, P_k^{(i)} \rangle \eqdef
\Esp{P_j^{(i)} (X_i) P_k^{(i)}(X_i) } = \int P_j^{(i)} (x_i)
P_k^{(i)}(x_i) f_{X_i}(x_i) \di x_i = c_j^{(i)} \delta_{jk}
 \label{eq3.3}
\end{equation} where $\langle \cdot,\cdot \rangle$ is the inner product
defined on the space associated with the probability measure
$\Pp_{X_i}(\di x_i)$, $\delta_{jk}$ is the Kronecker symbol with
$\delta_{jk}=1$ if $j=k$, otherwise $\delta_{jk}=0$ and $c_j^{(i)}$ is a
constant.  The univariate polynomial $P_{j}^{(i)}$ belongs to a specific
class according to the distribution of $X_i$.  For instance, if $X_i$ is
standard uniform (resp. Gaussian) random variable,
$\acc{P_j^{(i)}}_{j\ge0}$ are orthogonal Legendre
(resp. Hermite) polynomials.  Then the orthonormal univariate polynomials are
obtained by normalization:
\begin{equation} \Psi_j^{(i)}= P_j^{(i)} / \sqrt{c_j^{(i)}}
 \label{eq3.4}
\end{equation} Introducing the multi-indices $\ua= (\alpha_1 \enum
\alpha_M)$, a multivariate polynomial can be defined by tensor product as:
\begin{equation} \Psi_{\ua}(\vex) \eqdef \prod\limits_{i=1}^M
\Psi_{\alpha_i}^{(i)}(x_i)
 \label{eq3.5}
\end{equation} \citet{Soize2004} prove that the set of all multivariate
polynomials $\Psi_{\ua}$ in the input random vector $\veX$ forms a basis
of the Hilbert space of second order random variables:
\begin{equation} 
  Y = \sum\limits_{\ua \in {\Nn}^M} a_{\ua}
  \Psi_{\ua}(\veX)
 \label{eq3.6}
\end{equation} where $a_{\ua}$'s are the deterministic coefficients of
the representation.

In practice, the input random variables are usually not standardized,
therefore it is necessary to transform the input vector into a set of
standard variables.  We define the isoprobabilistic transform
$\veZ=\ct^{-1} (\veX)$ which is a unique mapping from the original
random space of $X_i$'s onto a standard space of $M$ basic independent
random variables $Z_i$'s.  As an example $Z_i$ may be a standard normal
random variable or a uniform variable over $[-1,1]$.

In engineering applications, only a finite number of terms can be
computed in Eq.(\ref{eq3.6}).  Accordingly, the truncated polynomial
chaos expansion of $Y$ can be represented as follows~\cite{Sudret2007}:
\begin{equation} Y=\cm(\ve{X})= \cm \prt{\ct(\ve{Z})} = \sum\limits_{\ua
\in \ca} a_{\ua}\Psi_{\ua}(\ve{Z})
\label{eq3.7}
\end{equation} in which $\ca$ is the set of multi-indices $\ua$'s
retained by the truncation scheme.

The application of PCE consists in choosing a suitable polynomial basis
and then computing the appropriate coefficients $a_{\ua}$'s.  To this
end, there exist several techniques including spectral projection
\cite{Lemaitre02, Matthies2005}, stochastic collocation method
\cite{Xiu2009a} or least square analysis (also called regression, see
\citep{BerveillerPMC04, Berveiller2006a}).  A review of these so-called
non-intrusive techniques is given in \citep{BlatmanThesis}. Recently,
the least-square approach has been extended to obtain {\em sparse}
expansions \citep{BlatmanPEM2010,BlatmanJCP2011}. This technique has
been applied to global sensitivity analysis in \cite{BlatmanRESS2010}.
The main results are now summarized.

First note that the orthonormality of the polynomial basis leads to the
following properties:
\begin{equation} \Esp{\Psi_{\ua}(\veZ)}=0 \qquad \text{and} \qquad
\Esp{\Psi_{\ua}(\veZ) \, \Psi_{\ub}(\veZ)}= \delta_{\ua \ub}
 \label{eq3.8}
\end{equation}
As a consequence the mean value of the model output $y$ is $\Esp{Y}=
a_0$ whereas the variance is the sum of the square of the other
coefficients:
\begin{equation}
 D=\Var{Y}= \Var{\sum\limits_{\ua \in \ca} a_{\ua}\Psi_{\ua}(\ve{Z})}= \sum\limits_{\begin{subarray}{c} \ua \in \ca \\ \ua \neq \bs{0} \end{subarray}} {a_{\ua}}^2 \, \Var{\Psi_{\ua}(\ve{Z})} 
 = \sum\limits_{\begin{subarray}{c} \ua \in \ca \\ \ua \neq \bs{0} \end{subarray}} {a_{\ua}}^2 
 \label{eq3.9}
\end{equation}
Making use of the unique orthonormality properties of the basis,
\citet{SudretCSM2006, SudretRESS2008b} proposed an original
post-processing of the PCE for performing global sensitivity analysis.  For any
subset variables $\iu=\acc{i_1 \enum i_s} \subset \acc{1 \enum M}$, one
defines the set of multivariate polynomials $\Psi_{\ua}$ which depends
only on $\iu$:
\begin{equation}
 \ca_{\iu}=\acc{ \ua \in \ca: \alpha_k \neq 0 \, \text{if and only if} \, k \in \iu  }
 \label{eq3.10}
\end{equation}
The $\ca_{\iu}$'s form a partition of $\ca$, thus the Sobol' decomposition of the truncated PCE in \eqrefe{eq3.7} may be written as follows:
\begin{equation}
 Y= a_0 + \sum\limits_{\begin{subarray}{c} \iu \subset \acc{1 \enum M} \\ \iu \neq \emptyset \end{subarray}} \cm_{\iu}(\veZ_{\iu})
 \label{eq3.11}
\end{equation}
where:
\begin{equation}
 \cm_{\iu}(\veZ_{\iu}) \eqdef \sum\limits_{\ua \in \ca_{\iu} } a_{\ua} \Psi_{\ua} (\veZ)
 \label{eq3.12}
\end{equation}
In other words the Sobol' decomposition is directly read from the PC
expansion.  Consequently, due to the orthonormality of PC basis, the
partial variance $D_{\iu}$ reads:
\begin{equation}
 D_{\iu}= \Var{ \cm_{\iu} (\veZ_{\iu})} = \sum\limits_{\ua \in \ca_{\iu}} a_{\ua}^2
 \label{eq3.13}
\end{equation}
As a cosequence the Sobol' indices at any order may be computed by a
mere combination of the squares of the coefficients.  As an
illustration, the first order PC-based Sobol' indices read:
\begin{equation}
 S_i = \sum\limits_{\ua \in \ca_{i}} a_{\ua}^2 / D, \quad \ca_{i}=\acc{ \ua \in \ca: \alpha_i > 0, \, \alpha_{j\neq i}=0 }
 \label{eq3.14}
\end{equation}
whereas the total PC-based Sobol' indices are:
\begin{equation}
 S_i^T = \sum\limits_{\ua \in \ca_{i}^T} a_{\ua}^2 / D, \quad \ca_{i}^T=\acc{ \ua \in \ca: \alpha_i > 0 }
\label{eq3.15}
\end{equation}

\section{Derivative of polynomial chaos expansions}
\label{sec04}

In this paper, we consider the combination of polynomial chaos
expansions with derivative-based global sensitivity analysis. On the one
hand, PCE are already known to provide accurate metamodels at reasonable
cost.  On the other hand, the derivative-based sensitivity measure
(DGSM) is effective for screening unimportant input factors.  The
combination of PCE and DGSM appears as a promising approach for
effective low-cost SA.  In fact, once the PCE metamodel is built, the
DGSM can be computed as a mere post-processing of the metamodel which
simply consists of polynomial functions.

As seen in Section \ref{sec02}, a DGSM is related to the expectation of
the square of the model derivative, which was denoted by $\nu_i$.  We
will express the model derivative in a way such that the expectation
operator can be easily computed, more precisely by projecting the
components of the gradient $\nabla \cm$ onto a PC expansion.

\subsection{Hermite polynomial chaos expansions}
\label{sec04.1} 
In this section we consider a numerical model $Y=\cm(\ve{X})$ where $Y$
is the scalar output and $\veX=\acc{ X_i \enum X_M}$ is the input vector
composed of $M$ \emph{independent Gaussian} variables $X_i \sim \cn
\prt{\mu_i,\sigma_i}$.  The isoprobabilistic transform reads:
\begin{equation} \veX= \ct(\veZ): \qquad X_i= \mu_i + \sigma_i Z_i
\end{equation} where $Z_i \sim \cn(0,1)$ are standard normal random
variables. The truncated PCE of $Y$ reads:
\begin{equation} Y=\cm(\ve{X})= \cm \prt{\ct(\ve{Z})} = \sum\limits_{\ua
\in \ca} a_{\ua} \Psi_{\ua}(\ve{Z})
\label{eq4.1.2}
\end{equation} in which $\ua=\acc{\alpha_1 \enum \alpha_M}$ is a
multi-index, $\ca$ is the set of indices $\ua$ in the truncated
expansion, $\Psi_{\ua}(\vez) = \prod\limits_{i=1}^M
\tilde{He}_{\alpha_i}(z_i)$ is the multivariate polynomial basis
obtained as the tensor product of univariate orthonormal Hermite
polynomials $\tilde{He}_{\alpha_i}(z_i)$ (see~\ref{appA}) and $a_{\ua}$
is the deterministic coefficient associated with $\Psi_{\ua}(\vez)$.

Since $\ct$ is a one-to-one mapping with $\dfrac{\partial z_i}{\partial
x_i} = \dfrac{1}{\sigma_i}$, the derivative-based sensitivity index
reads:
\begin{equation} \nu_i= \Esp{\prt{ \dfrac{\partial \cm}{\partial x_i}
(\ve{X}) }^2} = \Esp{ \prt{\dfrac{\partial \cm \circ \ct}{\partial z_i}
\dfrac{\partial z_i}{\partial x_i} }^2}= \dfrac{1}{{\sigma_i}^2}
\Esp{\prt{ \dfrac{\partial \cm \circ \ct }{\partial z_i}(\ve{Z}) }^2}
\end{equation}

The DGSM of $X_i$, in other words the corresponding upper bound to the
total Sobol' index $S_i^T$, is computed according to \eqrefe{eq14}:
\begin{equation} 
S_i^{DGSM}= \sigma_i^2 \dfrac{\nu_i}{D} =\dfrac{1}{D}
\Esp{\prt{ \dfrac{\partial \cm \circ \ct }{\partial z_i}(\ve{Z}) }^2} =
\dfrac{1}{D} \Esp{\prt{ \dfrac{\partial}{\partial z_i} \sum\limits_{\ua
\in \ca} a_{\ua} \Psi_{\ua} (\ve{Z}) }^2}
\label{eq26}
\end{equation} 
in which $D=\Var{Y}=\sum\limits_{\ua \in \ca,\,\ua \neq
\ve{0}} a_{\ua}^2$.  This requires computing the partial derivatives of
polynomial functions of the form $\cm_{\ca}(\vez)= \sum\limits_{\ua \in
\ca} a_{\ua} \Psi_{\ua}(\vez)$. One can prove that the derivatives
$\tilde{He}_n^{'}(z) = \dfrac{\di \tilde{He}_n}{\di z}(z)$ read (see
\ref{appA}):
 \begin{equation} 
\tilde{He}_n^{'}(z)= \sqrt{n}\, \tilde{He}_{n-1}(z)
 \end{equation} 
Therefore the derivative of the multivariate orthonormal
Hermite polynomial $\Psi_{\ua}(\vez) = \prod\limits_{i=1}^M
\tilde{He}_{\alpha_i} (z_i) $ with respect to $z_i$ reads:
 \begin{equation} 
\dfrac{\partial \Psi_{\ua}}{\partial z_i} (\vez) =
\prod\limits_{\substack{j=1\\j \neq i}}^M \tilde{He}_{\alpha_j} (z_j) \,
\sqrt{\alpha_i} \tilde{He}_{\alpha_i-1} (z_i)
\end{equation} 
provided that $\alpha_i >0 $, and $\dfrac{\partial \Psi_{\ua}}{\partial
  z_i} (\vez) =0$ otherwise.  Then the derivative of a Hermite PCE with
respect to $z_i$ is  given the following expression:
 \begin{equation} \dfrac{\partial \cm_{\ca}}{\partial z_i} (\vez)=
\sum\limits_{\ua \in \ca^{(i)}} \sqrt{\alpha_i} \, a_{\ua}
\Psi_{\ua_i^{'}}
\end{equation} 
in which $\ca^{(i)}= \{ \ua \in \ca, \alpha_i >0 \}$ is the set of
multi-indices $\ua$ having a non-zero $i^{th}$ component and $\ua_i^{'}=
\{ \alpha_1 \enum \alpha_{i}-1 \enum \alpha_M \}$ is the index vector
derived from $\ua$ by subtracting 1 from $\alpha_i$.  The expectation of
the squared derivative in \eqrefe{eq26} is reformulated as:
 \begin{equation} 
   \Esp{ \prt{\dfrac{\partial \cm_{\ca}}{\partial z_i}
       (\veZ)}^2 } = \Esp{ \sum\limits_{\ua \in \ca^{(i)}} \sum\limits_{\ub \in
       \ca^{(i)}} \sqrt{\alpha_i\, \beta_i} \, a_{\ua} a_{\ub} \,
     \Psi_{{\ua_i}^{'}} \Psi_{{\ub}_i^{'}} }
\end{equation} 
Due to the linearity of the expectation operator, the above equation
requires computing $\Esp{\Psi_{{\ua_i}^{'}} \Psi_{{\ub}_i^{'}}}$.  Note
that the orthonormality of the polynomial basis leads to
$\Esp{\Psi_{{\ua_i}^{'}} \Psi_{{\ub}_i^{'}}} = \delta_{\ua\ub}$ where
$\delta_{\ua\ub}$ is the Kronecker symbol. Thus one has:
\begin{equation} 
\Esp{\prt{\dfrac{\partial \cm_{\ca}}{\partial z_i}
(\veZ)}^2} = \sum\limits_{\ua \in \ca^{(i)}} \alpha_i \, a_{\ua}^2
\end{equation}
As a consequence, in case of a Hermite PCE the DGSM can be given the
following {\em analytical expression}:
\begin{equation} \hat{S}_i^{DGSM}= \dfrac{1}{D} \sum\limits_{\ua \in
\ca^{(i)}} \alpha_i \,a_{\ua}^2 = \dfrac{\sum\limits_{\ua \in \ca^{(i)}}
\alpha_i \, a_{\ua}^2}{\sum\limits_{\ua \in \ca ,\,\ua \neq \ve{0}}
a_{\ua}^2}
\end{equation} Note that the total Sobol' indices $S^T_i$ can be
obtained directly from the PCE by $ \dsp{\hat{S}_i^{T}= \sum\limits_{\ua
\in \ca^{(i)}} a_{\ua}^2 / \sum\limits_{\ua \in \ca ,\,\ua \neq \ve{0}}
a_{\ua}^2 }$ as shown in~\eqrefe{eq3.15}. With integer indices
$\alpha_i>0$, it is clear that the inequality $S_i^T \leq S_i^{DGSM}$ is
always true by construction.

\subsection{Legendre polynomial chaos expansions}
\label{sec04.2} 
Consider now a computational model $Y=\cm(\ve{X})$ where the input
vector $\veX$ contains $M$ independent uniform random variables $X_i
\sim \cu [a_i,\,b_i]$.  We first use an isoprobabilistic transform to
convert the input factors into normalized variables $\veZ=\acc{Z_i \enum
  Z_M}$:
\begin{equation} 
  \veX= \ct(\veZ): \qquad X_i= \dfrac{b_i + a_i}{2} +
  \dfrac{b_i - a_i}{2} Z_i
\end{equation} 
where $Z_i \sim \cu[-1,1]$ are uniform random variables.  The Legendre
PCE has the form of the expansion in \eqrefe{eq4.1.2}, except that
$\Psi_{\ua}(\ve{z})= \prod\limits_{i=1}^M \tilde{Le}_{\alpha_i}(z_i)$ is
now the multivariate polynomial basis made of univariate orthonormal
Legendre polynomials $\tilde{Le}_{\alpha_i}(z_i)$ (see~\ref{appB}).
Again, since $\ct$ is a one-to-one linear mapping with $\dfrac{\partial
  z_i}{\partial x_i} = \dfrac{2}{b_i-a_i}$ the derivative-based
sensitivity index reads:
\begin{equation}
  \nu_i  = \Esp{\prt{ \dfrac{\partial \cm}{\partial x_i} (\ve{X}) }^2}
   = \dfrac{4}{(b_i - a_i)^2} \, \Esp{\prt{ \dfrac{\partial \cm \circ \ct
        }{\partial z_i} (\ve{Z})}^2}
\end{equation}
Similarly to \eqrefe{eq26}, the upper bound DGSM to the total Sobol'
index $S_i^T$ is computed from \eqrefe{eq13} as:
\begin{equation}
  \begin{split}
    S_i^{DGSM} &= \dfrac{(b_i-a_i)^2}{\pi^2} \dfrac{\nu_i}{D} =
    \dfrac{4}{\pi^2 D} \, \Esp{\prt{ \dfrac{\partial \cm \circ \ct
        }{\partial z_i} (\ve{Z})}^2} \\
    &= \dfrac{4}{\pi^2 D} \,\Esp{ {\prt{
          \dfrac{\partial} {\partial z_i} \sum\limits_{\ua \in \ca}
          a_{\ua} \, \Psi_{\ua} (\veZ)} } ^2}
  \end{split}\end{equation}

Thus the derivative of univariate and multivariate Legendre polynomials
are required. Denoting by $\tilde{Le}_i^{'}(z)\eqdef \dfrac{\di
  \tilde{Le}(z)}{\di z}$, one shows in \ref{appB} that:
\begin{equation}
  \acc{ \tilde{Le}_1^{'}(z) \enum \tilde{Le}_n^{'}(z) }\tr =
  \mat{C}^{\cl e}
  \cdot \{ \tilde{Le}_0(z) \enum \tilde{Le}_{n-1}(z) \}\tr
\end{equation}
in which $\mat{C}^{\cl e}$ is a constant matrix whose $i^{th}$ row
contains the coordinates of the derivative of $\tilde{Le}_i(z)$ onto a
basis made of lower-degree polynomials $\acc{\tilde{Le}_j(z),\, j=0
  \enum i-1}$.  In other words, $\tilde{Le}_i^{'}(z) =
\sum\limits_{j=1}^{i} C_{ij}^{\cl e} \tilde{Le}_{j-1} (z)$.
%
Using this notation, the derivative of the multivariate orthonormal Legendre polynomials $\Psi_{\ua}(\vez) = \prod\limits_{i=1}^M \tilde{Le}_{\alpha_i} (z_i) $ with respect to $z_i$ reads:
 \begin{equation}
   \dfrac{\partial \Psi_{\ua}}{\partial z_i} (\vez) =
   \prod\limits_{\substack{j=1\\j \neq i}}^M \tilde{Le}_{\alpha_j} (z_j)
   \prt{ \sum\limits_{l=1}^{\alpha_i} C_{\alpha_i l}^{\cl e} \tilde{Le}_{l-1}
     (z_i) }
     \label{eq4.13}
\end{equation}
For a given $\ua = \acc{ \alpha_1 \enum \alpha_M }$ let us define by $\ua_i^r$ the index vector having the $i^{th}$ component equal to $r$:
\begin{equation}
 \ua_i^r = \acc{ \alpha_1 \enum \overbrace{r}^{i^{th} position} \enum \alpha_M }
\end{equation}
Using this notation \eqrefe{eq4.13} rewrites as follows:
\begin{equation}
   \dsp{\dfrac{\partial \Psi_{\ua}}{\partial z_i} (\vez) =
      \sum\limits_{l=1}^{\alpha_i} C_{\alpha_i l}^{\cl e} \Psi_{\ua_i^{l-1}}} 
\end{equation}
Denote by $\ca^{(i)}$ the set of $\ua$ having a non-zero index $\alpha_i$, \ie 
$\ca^{(i)}= \{ \ua \in \ca, \alpha_i >0 \}$.
The derivative of a Legendre PCE with respect to $z_i$ then reads:
 \begin{equation}
\label{eq:059}
 \dfrac{\partial \cm_{\ca}}{\partial z_i} (\vez)= \sum\limits_{\ua \in
   \ca^{(i)}} a_{\ua} \dfrac{\partial \Psi_{\ua}}{\partial z_i} (\vez)= \sum\limits_{\ua \in
   \ca^{(i)}} \sum\limits_{l=1}^{\alpha_i} a_{\ua} \, C_{\alpha_i l}^{\cl} \, 
 \Psi_{\ua_i^{l-1}} (\vez) 
\end{equation}
Denote by $\cb^{(i)}$ the set of multi-indices $\ub$ representing the ensemble of multivariate 
polynomials generated by differentiating the linear combination of polynomials $\acc{\Psi_{\ua}(\vez),\, \ua \in \ca^{(i)} }$. $\cb^{(i)}$ is obtained as:
\begin{equation}
\cb^{(i)} = \acc{\ub = \ua + (k-\alpha_i) \cdot \ve{e}_i,\,\ua \in \ca^{(i)}, \,k=0 \enu \alpha_i-1}
\end{equation}
where:
\begin{equation}
 \ve{e}_i= (0, \dots, 0, \overbrace{1}^{i^{th} \text{pos.}},0 \dots,0)
\end{equation}
The derivative of Legendre PCE rewrites:
 \begin{equation}
 \dfrac{\partial \cm_{\ca}}{\partial z_i} (\vez)=  
 \sum_{\ub \in \cb^{(i)}} b_{\ub} \, \Psi_{\ub}(\vez)
\end{equation}
in which the coefficient $b_{\ub}$ is obtained from Eq.(\ref{eq:059}).
Since the polynomials $\Psi_{\ub}$ are also orthonormal, one obtains:
\begin{equation}
\Esp{\prt{\dfrac{\partial \cm_{\ca}}{\partial z_i} (\veZ)}^2}=
\sum_{\ub \in \cb^{(i)}}  b_{\ub}^2
\end{equation}
Finally, the DGSMs read:
\begin{equation}
\hat{S}_i^{DGSM} = \dfrac{4}{\pi^2} \dfrac{\sum\limits_{\ub \in
    \cb^{(i)}} b_{\ub}^2} {\sum\limits_{\ua \in \ca ,\,\ua \neq \ve{0}} a_{\ua}^2}
    \label{eq42}
\end{equation}
\subsection{General case}
\label{sec04.3}
Consider now the general case where the input vector $\veX$ contains $M$
independent random variables with different prescribed probability
distribution functions, \ie Gaussian, uniform or others. Such a problem can be
addressed using generalized polynomial chaos expansions \cite{Xiu2002}.
As the above derivations for Hermite and Legendre polynomials are valid
componentwise, they remain identical when dealing with generalized
expansions. Only the proper matrix yielding the derivative of the
univariate polynomials in the same univariate orthonormal basis is needed,
see Appendix~A for Hermite polynomials and Appendix~B for Legendre
polynomials. The derivation for Laguerre polynomials is also given in
Appendix~C for the sake of completeness.

\section{Application examples}
\label{sec05}

\subsection{Morris function}
\label{sec05.1}
We first consider the Morris function that is widely used in the literature for sensitivity analysis~\cite{Morris1991, Lamboni2013}. 
This function reads:
\begin{equation}
y=\beta_o + \sum_{i=1}^{20}\beta_i \,\omega_i + \sum_{i<j}^{20}
\beta_{ij}\, \omega_i \, \omega_j + \sum_{i<j<l}^{20} \beta_{ijl}\, \omega_i
\, \omega_j \, \omega_l+ \beta_{1234} \,\omega_1 \, \omega_2 \, \omega_3 \, \omega_4
\end{equation}
in which:
\begin{itemize}
 \item $\omega_i=2\prt{X_i-1/2}$ except for $i=3,5,7$ where $\omega_i=2\prt{1.2\dfrac{X_i}{X_i+1} -\dfrac{1}{2}}$,
 \item the input vector $\veX=\acc{X_1 \enu X_{20}}$ contains 20 uniform random variables $\acc{X_i \sim \cu[0,1],\, i=1\enum 20}$,
 \item $\beta_i=20$ for $i=1,2 \enu 10$,
 \item $\beta_{ij}=-15$ for $i,j=1,2 \enu 6, \, i<j$,
 \item $\beta_{ijl}=-10$ for $i,j,l=1,2 \enu 5, \, i<j<l$,
 \item $\beta_{1234}=5$,
 \item the remaining first and second order coefficients are defined by $\beta_i=(-1)^i$, $\beta_0=0$ and $\beta_{ij}=(-1)^{i+j}$,
 \item and the remaining third order coefficients are set to 0.
\end{itemize}

First, a PCE is built using the Least Angle Regression technique based
on a Latin Hypercube experimental design of size $N=500$.  Then the PCE
is post-processed to obtain the total Sobol' indices and the upper-bound
derivative-based sensitivity measures (DGSMs) using \eqrefe{eq3.15} and
\eqrefe{eq42}, respectively.  The procedure is replicated 100 times in
order to provide the 95\% confidence interval of the resulting
sensitivity indices.

As a reference, the total Sobol' indices are computed by Monte Carlo
simulation as described in Section~\ref{sec02.1} using the
\texttt{sensitivity} package in \texttt{R}~\cite{Rsensitivity}.  One
samples two experimental designs of size $N=5,000$ denoted respectively
by $A$ and $B$ then computes the corresponding output vectors $Y_A$ and
$Y_B$.  To estimate the total sensitivity index $S_i^T$ with respect to
random variable $X_i$, one replaces the entire $i^{th}$ column in sample
$A$ (which contains the samples of $X_i$) by the $i^{th}$ column in
sample $B$ to obtain a new experimental design denoted by $C_i$.  Then
the output $Y_{C_i}$ is computed from the input $C_i$. The
variance-based $S_i^T$ is obtained by means of $Y_A$, $Y_B$ and
$Y_{C_i}$ using the \texttt{sobol2007} function~\cite{Rsensitivity,
  Saltelli2010}.  The total number of model evaluations required by the
MCS approach is $5,000 \times (2+20)=110,000$. After 100 replications we
also obtain the 95\% confidence interval of the sensitivity indices.

\begin{figure}[!ht]
\centering
 \includegraphics[width=0.7\linewidth]{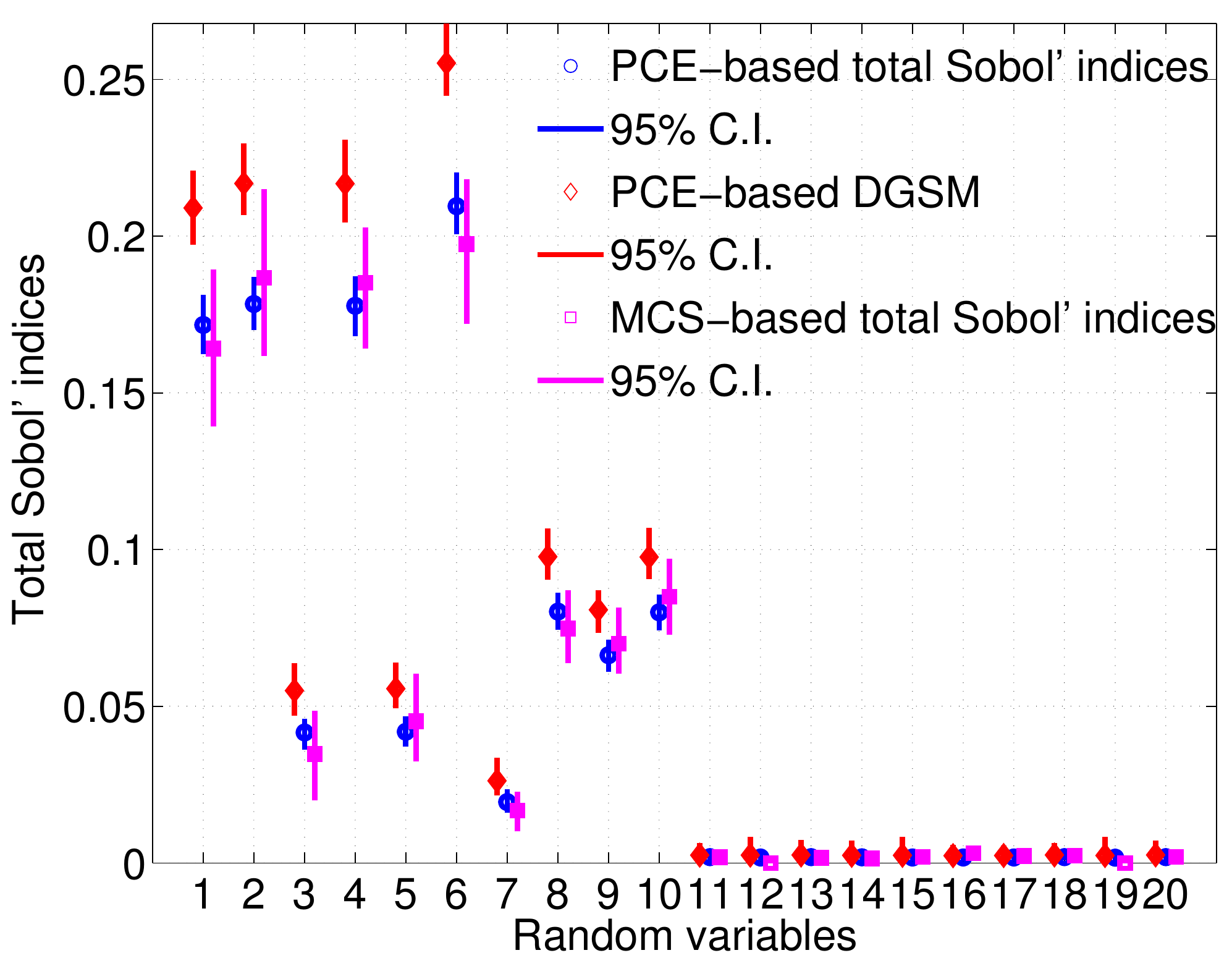}
 \caption{Morris function: PCE-based \emph{vs.} MCS-based sensitivity measures}
 \label{fig1}
\end{figure}

For each input variable, \figref{fig1} depicts the total sensitivity
indices computed by MCS and PCE approaches, and the DGSM derived from
PCE as well as their corresponding 95\% confidence intervals (the bounds
of the latter being obtained from the 100 replications). \figref{fig1}
shows that PCE derivative-based sensitivity measures and total Sobol'
indices for parameters $X_{11} \enum X_{20}$ are both close to zero, \ie
$X_{11} \enum X_{20}$ are unimportant factors.  This is consistent with
the MCS-based sensitivity measures. Using a sample set of size $500$,
the PCE-based approach can detect the non-significant input parameters
with acceptable accuracy compared to MCS which requires $110,000$ model
runs. In addition, the PCE-based total Sobol' indices for the remaining
parameters vary in a significantly narrower confidence intervals, \ie
are more reliable, than the MCS-based sensitivity indices.  Finally, the
obtained total Sobol' indices are always smaller than the DGSMs.  The
less significant the parameter, the closer DGSM gets to the total Sobol'
index.

\subsection{Oakley \& O'Hagan function}
The second numerical example is the Oakley \& O'Hagan function~\cite{Sobol2009,Oakley2004} which reads:
\begin{equation}
 f(\ve{X})= \ve{a}_1^T \ve{X} +\ve{a}_2^T \cos (\ve{X}) +\ve{a}_3^T
 \sin(\ve{X}) +\ve{X}^T \mat{M} \ve{X} 
\end{equation}
in which the input vector $\veX=\acc{X_1 \enu X_{15}}$ consists of 15 independent standard normal random 
variables $\acc{ X_i \sim \cn(0,1),\, i=1 \enum 15}$. The $15 \times 1$ vectors $\ve{a}_j, \, j=1,2,3$ 
and the $15 \times 15 $ matrix $\mat{M}$ are provided at \url{www.sheffield.ac.uk/st1jeo}.

Given the complexity of the function, the PCE-based approach is run with a Latin Hypercube experimental design of size $N=600$. 
The size of a single sample set for the MCS approach is $N=10,000$ resulting in  $10,000 \times (2+15)= 170,000$ model runs. 
The procedure is similar as in Section~\ref{sec05.1}.

\begin{figure}[!ht]
\centering
 \includegraphics[width=0.7\linewidth]{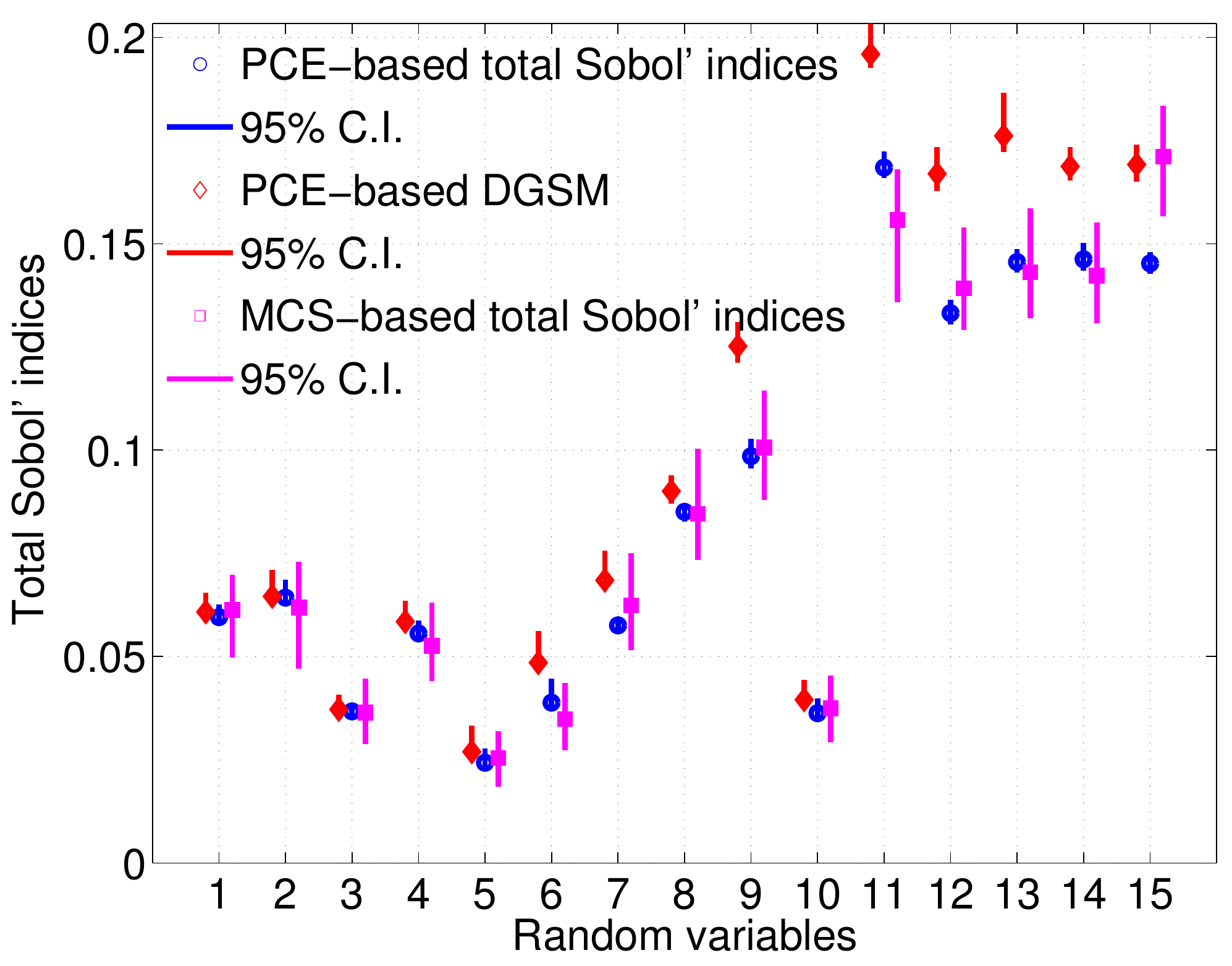}
 \caption{Oakley \& O'Hagan function: PCE-based \emph{vs.} MCS-based sensitivity measures}
 \label{fig2}
\end{figure}

\figref{fig2} shows that the PCE-based approach using only $600$ model
evaluations can estimate the least significant parameters
$X_3,\,X_5,\,X_6,\,X_{10}$ with {less uncertainty} compared to the MCS
approach that uses a huge number of model runs.  As already observed in
\figref{fig1}, the DGSMs which are the upper bounds of the total Sobol'
indices, are close to the latter for the parameters whose total Sobol'
indices are smaller than 5\%.  For the remaining parameters which are
more influential, the differences between the total Sobol' indices and
their upper bounds become larger.

\section{Conclusions}

In practical problems, the systems of interest usually contain numerous random input 
factors which might lead to large uncertainty in the system output and high computational cost. 
Therefore, it is important to quantify the important and unimportant factors according to their contributions to the output uncertainty.
However, the commonly used Monte Carlo simulation (MCS) approach is usually computationally prohibitive.

In this paper, we combined the polynomial chaos expansions (PCE)
metamodelling with a derivative-based sensitivity analysis technique.
Polynomial chaos expansions are effective surrogate models for global
sensitivity analysis.  The very nature of the orthogonal expansions
reduces the computation of (total) Sobol' indices to a mere
post-processing of the PC coefficients. Similarly, DGSMs can be computed
by a straightforward post-processing, \ie without requiring additional
model runs.  One only needs to differentiate the multivariate
polynomials, which in the end reduces to differentiating univariate
polynomial functions. Expressions were given for the classical Hermite,
Legendre and Laguerre polynomials. In order to carry out the computation
efficiently the derivative polynomials shall be represented onto the
orthonormal basis of the same family, which can be do once and for all.
Then computing the DGSM reduces to computing weighted sums of the
polynomial chaos coefficients.  The technique is illustrated on two
well-known benchmark functions. By comparing with Monte Carlo
simulation, the PCE approach is shown to provide sensitivity indices
with smaller uncertainty at a computational cost that is several orders
of magnitude smaller.

\bibliographystyle{model3-num-names}
\bibliography{IBK-Sudret}

\medskip
\medskip
\appendix

As seen in Section \ref{sec04}, the computation of polynomial chaos
expansions derivative-based global sensitivity measures (PCE-DGSMs)
consists of two steps.  The first step is to represent the derivative of
the PCE in terms of orthonormal polynomials from the same families.  It
essentially requires  to construct the matrices of coefficients $\mat{C}$
that are used for differentiating the classical orthonormal polynomials.
The second step is to post-process this ``PCE'' of the derivative. A
general solution to compute the mean squared derivative using the
coefficients matrices $\mat{C}$ was presented in Section \ref{sec04.3}.

\section{Hermite polynomial chaos expansions}
\label{appA}

The classical Hermite polynomials $\acc{He_{n}, n\in \Nn }$, where $n$
determines the degree of the polynomial, are defined on the set of real
numbers$\Rr$ so as to be orthogonal with respect to the Gaussian
probability measure and associated inner product:
\begin{equation}
 \langle He_m,\, He_n \rangle \eqdef \int_{\Rr} He_m (z) He_n (z) \dfrac{e^{-z^2/2}}{\sqrt{2\pi}} \, \di z = n!\, \delta_{mn}
 \label{eqA1}
\end{equation}
The  Hermite polynomials satisfy the following differential equation \cite[Chap. 22]{Abramovitz1965}
\begin{equation}
\frac{\di}{\di z} He_n(z) = n He_{n-1}(z)
\label{eqA2}
\end{equation}
From \eqrefe{eqA1} the norm of Hermite polynomials reads:
\begin{equation}
 \langle He_n,\, He_n \rangle = n!
 \label{eqA3}
\end{equation}
so that the \emph{orthonormal Hermite polynomials} are defined by:
   \begin{equation}
  \tilde{He}_n(z)=\frac{1}{\sqrt{n!}} He_n(z)
  \label{eqA4}
  \end{equation}
Substituting for \eqrefe{eqA4} in \eqrefe{eqA2}, one gets the derivative of 
orthonormal Hermite polynomial $\tilde{He}_n^{'}(z) \eqdef \dfrac{\di \tilde{He}(z)}{\di z}$:
\begin{equation}
 \tilde{He}_n^{'}(z)= \sqrt{n}\, \tilde{He}_{n-1}(z)
\label{eqA5}
 \end{equation}
For computational purposes the following matrix notation is introduced:
\begin{equation}
 \acc{ \tilde{He}_1^{'}(z) \enum \tilde{He}_n^{'}(z) }\tr = \mat{C}^{\ch}
 \cdot \{ \tilde{He}_0(z) \enum \tilde{He}_{n-1}(z) \}\tr
 \label{eqA6}
\end{equation}
which allows one to cast the derivative of the orthonormal Hermite polynomials in the initial basis. 
From \eqrefe{eqA5}, $\mat{C}^{\ch}$ is obviously diagonal:
 \begin{equation}
  \mat{C}_{i,j}^{\ch}=\sqrt{i} \, \delta_{ij}
 \end{equation}
\section{Legendre polynomial chaos expansions}

\label{appB}

The classical Legendre polynomials $\acc{Le_{n}, n\in \Nn }$ are defined
over $[-1,1]$ so as to be orthogonal with respect to the uniform
probability measure and associated inner product:
\begin{equation}
 \langle Le_m,\, Le_n \rangle \eqdef \int_{-1}^1 Le_m (z) Le_n (z) \dfrac{\di z} {2} = \dfrac{1}{2\, n +1}\, \delta_{mn}
 \label{eqB1}
\end{equation}
They satisfy the following differential equation \citep[Chap.
22]{Abramovitz1965}
\begin{equation}
\frac{\mathrm{d}}{\mathrm{d} z} \bra{Le_{n+1}(z)-Le_{n-1}(z)} = (2n+1)
Le_n(z)
\label{eqB2}
\end{equation}
Using the notation ${Le}_n^{'}(z) \eqdef \dfrac{\di Le_n(z)}{\di z}$ one
can transform \eqrefe{eqB2} into the equation:
\begin{equation}
 \begin{aligned}
    Le_{n+1}^{'}(z)  &=(2n+1)Le_n(z)+Le_{n-1}^{'}(z)\\
   & =(2n+1)Le_n(z)+(2(n-2)+1)Le_{n-2}(z) +Le_{n-3}^{'}(z) \\
   & = \cdots
 \end{aligned}
 \label{eqB3}
\end{equation}
From \eqrefe{eqB1}, the norm of Legendre polynomials reads:
\begin{equation}
 \langle Le_n,\, Le_n \rangle = \dfrac{1}{2\,n+1}
 \label{eqB4}
\end{equation}
so that the \emph{orthonormal Legendre polynomials} read:
 \begin{equation}
  \tilde{Le}_n(z) = \sqrt{2n+1} \, Le_n(z)
  \label{eqB5}
\end{equation}
Substituting for \eqrefe{eqB5} in \eqrefe{eqB3} one obtains:
\begin{equation}
\begin{split}
  \tilde{Le}_{n+1}^{'}(z)=\sqrt{2n+3} \, &\Big[ \sqrt{2n+1}
    \, \tilde{Le}_n(z) + \sqrt{2(n-2)+1} \, \tilde{Le}_{n-2}(z)\\ &
    +\sqrt{2(n-4)+1} \, \tilde{Le}_{n-4}(z) + \dots \Big]
\end{split}
\label{eqB6}
\end{equation}
Introducing the matrix notation:
\begin{equation}
 \acc{ \tilde{Le}_1^{'}(z) \enum \tilde{Le}_n^{'}(z) }\tr = \mat{C}^{\cl
 e}
 \cdot \{ \tilde{Le}_0(z) \enum \tilde{Le}_{n-1}(z) \}\tr
 \label{eqB7}
\end{equation}
the matrix $\mat{C}^{\cl e}$ reads:
  \begin{equation}
\mat{C}^{\cl e} =
 \begin{bmatrix}
   \sqrt{3} & 0 & 0 & 0 & \ldots \\
   0 & \sqrt{5}\sqrt{3} & 0 & 0 & \ldots  \\
   \sqrt{7}\cdot1 & 0 & \sqrt{7}\sqrt{5} & 0 & \ldots   \\
   \vdots \\
   0 & \sqrt{4p+1} \sqrt{3} & 0 & \sqrt{4p+1}\sqrt{7} & \ldots &
   \sqrt{4n+1} \sqrt{4n-1} \\
\end{bmatrix}
 \end{equation}
 when $n=2p$ is even and
  \begin{equation}
\mat{C}^{\cl e}  =
 \begin{bmatrix}
   \sqrt{3} & 0 & 0 & 0 & \ldots \\
   0 & \sqrt{5}\sqrt{3} & 0 & 0 & \ldots  \\
   \sqrt{7}\cdot1 & 0 & \sqrt{7}\sqrt{5} & 0 & \ldots   \\
   \vdots \\
   \sqrt{4p+3}\cdot1 & 0 & \sqrt{4p+3}\sqrt{5} & 0 & \ldots & 0&
   \sqrt{4p+3}\sqrt{4p+1}
\end{bmatrix}
 \end{equation}
 when $n=2p+1$ is odd.

%
\section{Generalized Laguerre polynomial chaos expansions}

\label{appC}
Consider a model $Y=\cm(\ve{X})$ where the input vector $\veX$ contains
$M$ independent random variables with Gamma distribution $X_i \sim
\Gamma(\alpha_i,\beta_i), \, (\alpha_i,\beta_i >0)$ with prescribed
probability density functions:
\begin{equation}
 f_{X_i}(x_i)={\beta_i}^{\alpha_i} \, \dfrac{1}{\Gamma(\alpha_i)} x^{\alpha_i-1} e^{-\beta_i x_i}
 \label{eqC1}
\end{equation}
where $\Gamma(\cdot)$ is the Gamma function.
We first use an isoprobabilistic transform to convert the input factors
into a random vector $\veZ=\acc{Z_i \enum Z_M}$ as follows:
\begin{equation}
 Z_i=\beta_i \, X_i
 \label{eqC2}
\end{equation}
One can prove that:
\begin{equation}
 f_{Z_i}(z_i)= \abs{\dfrac{\di x_i}{\di z_i} } f_{X_i}(x_i)=\dfrac{1}{\Gamma(\alpha)} {z_i}^{\alpha-1} e^{-z_i}
 \label{eqC3}
\end{equation}
which means $Z_i \sim \Gamma(\alpha_i,1)$.

By definition, the generalized Laguerre polynomials
$\acc{L_n^{(\alpha-1)} (z),\, n \in \Nn }$, where $n$ is the degree of
the polynomial, are orthogonal with respect to the weight function
$w(z)= z^{\alpha-1} e^{-z}$ over $(0,\infty)$:
\begin{equation}
  \langle L_n^{(\alpha-1)} (z) , L_m^{(\alpha-1)} (z) \rangle \eqdef
  \int\limits_0^{+\infty} z^{\alpha-1} e^{-z} L_n^{(\alpha-1)} (z)
  L_m^{(\alpha-1)} (z) \di z = \dfrac{\Gamma(n+\alpha)}{n!} \delta_{mn} 
 \label{eqC4}
\end{equation}
The derivative of $L_n^{(\alpha-1)} $ reads:
\begin{equation}
 L_n^{'(\alpha-1)}(z) = - \sum_{k=0}^{n-1} L_k^{(\alpha-1)}(z)
 \label{eqC5}
\end{equation}
Recall that one obtains the Gamma distribution by scaling the weight
function $w(z)$ by $1/\Gamma(\alpha)$.  Therefore in the context of PCE,
we use the generalized Laguerre polynomials functions orthonormalized as
follows:
\begin{equation}
 \tilde{L}_n^{(\alpha-1)} (z) = \sqrt{
   \dfrac{n!\Gamma(\alpha)}{\Gamma(n+\alpha)} } L_n^{(\alpha-1)} (z)
 =\sqrt{n\, B(n,\alpha)} \, L_n^{(\alpha-1)} (z)
 \label{eqC6}
\end{equation}
where $B(x,y) = \frac{\Gamma(x)\Gamma(y)}{\Gamma(x+y)}$ is the beta
function. Substituting for \eqrefe{eqC6} in \eqrefe{eqC5} one obtains:
\begin{equation}
  \tilde{L}_n^{'(\alpha-1)}(z)= - \sum_{k=0}^{n-1} \sqrt{
    \dfrac{\Gamma(k+\alpha+1) \, n!}{ \Gamma(n+\alpha+1)\, k!} }
  \tilde{L}_k^{(\alpha-1)}(z) =
  -\sum_{k=1}^n \sqrt{\frac{B(n+1,\alpha)}{B(k,\alpha)}}
  \tilde{L}_{k-1}^{(\alpha-1)}(z)
 \label{eqC7}
\end{equation}
Introducing the matrix notation:
\begin{equation}
 \acc{ \tilde{L}_1^{'}(z) \enum \tilde{L}_n^{'}(z) }\tr = \mat{C}^{\cl a}
 \cdot \{ \tilde{L}_0(z) \enum \tilde{L}_{n-1}(z) \}\tr
 \label{eqC7-b}
\end{equation}
the constant matrix $\mat{C}^{\cl a }$ is a lower
triangular matrix whose generic term reads:
\begin{equation}
 \mat{C}_{i,j}^{\cl a} = - \sqrt{ \frac{B(i+1,\alpha)}{B(j,\alpha)} } 
\end{equation}

\label{lastpage}









\end{document}